\documentclass{comnet}

\usepackage{amsmath} 
\usepackage{amsthm} 
\usepackage{subfig} 

\usepackage{graphicx}
\usepackage{dcolumn}
\usepackage{bm}
\usepackage{color}
\usepackage{txfonts}
\usepackage{comment}
\usepackage{longtable}
\usepackage{url}
\usepackage{changepage}
\usepackage{rotating}

\definecolor{dkgreen}{rgb}{0,0.6,0}
\definecolor{gray}{rgb}{0.5,0.5,0.5}
\definecolor{mauve}{rgb}{0.58,0,0.82}

\begin{document}

\title{Mesoscale analyses of fungal networks as an approach for quantifying phenotypic traits}

\shorttitle{Mesoscale analyses of fungal networks} 
\shortauthorlist{S.\,H. Lee, M.\,D. Fricker, and M.\,A. Porter} 

\author{
\name{Sang Hoon Lee$^*$}
\address{School of Physics, Korea Institute for Advanced Study, Seoul 02455, Korea}
\address{Integrated Energy Center for Fostering Global Creative Researcher (BK 21 plus) and Department of Energy Science,
Sungkyunkwan University, Suwon 16419, Korea}
\address{Oxford Centre for Industrial and Applied Mathematics, Mathematical Institute, University of Oxford, Oxford OX2 6GG, United Kingdom\email{$^*$Corresponding author: lshlj82@kias.re.kr}}
\name{Mark D. Fricker}
\address{Department of Plant Sciences, University of Oxford, South Parks Road, Oxford OX1 3RB, United Kingdom}
\address{CABDyN Complexity Centre, University of Oxford, Oxford OX1 1HP, United Kingdom 
}
\and
\name{Mason A. Porter}
\address{Oxford Centre for Industrial and Applied Mathematics, Mathematical Institute, University of Oxford, Oxford OX2 6GG, United Kingdom}
\address{CABDyN Complexity Centre, University of Oxford, Oxford OX1 1HP, United Kingdom}}

\maketitle

\begin{abstract}
{We investigate the application of mesoscopic response functions (MRFs) to characterize a large set of networks of fungi and slime moulds grown under a wide variety of different experimental treatments, including inter-species competition and attack by fungivores. We construct `structural networks' by estimating cord conductances (which yield edge weights) from the experimental data, and we construct `functional networks' by calculating edge weights based on how much nutrient traffic is predicted to occur along each edge.  Both types of networks have the same topology, and we compute MRFs for both families of networks to illustrate two different ways of constructing taxonomies to group the networks into clusters of related fungi and slime moulds.  Although both network taxonomies generate intuitively sensible groupings of networks across species, treatments, and laboratories, we find that clustering using the functional-network measure appears to give groups with lower intra-group variation in species or treatments. We argue that MRFs provide a useful quantitative analysis of network behaviour that can (1) help summarize an expanding set of increasingly complex biological networks and (2) help extract information that captures subtle changes in intra-specific and inter-specific phenotypic traits that are integral to a mechanistic understanding of fungal behaviour and ecology.
As an accompaniment to our paper, we also make a large data set of fungal networks available in the public domain.
}

{fungal networks; transport networks; community structure; mesoscopic response functions; taxonomies; core--periphery structure; structural networks; functional networks, trait analysis}


\end{abstract}



\section{Introduction}
\label{sec:introduction}

Fungi are unusual multi-cellular macroscopic organisms: their entire form is a living network of interconnected microscopic tubular cells (called `hyphae') that can branch, fuse, or aggregate to form larger, visible structures (called `cords').
The resulting mycelial network has to transport nutrients from sites of acquisition to the growing tips to fuel further exploration for resources that exist with an unknown distribution in a fluctuating, patchy, and competitive environment~\cite{Heaton2012}. Mycelial networks also provide food for small grazing invertebrates, and they thus suffer continuous attack and damage~\cite{Crowther2012}. This highlights their essential roles --- including decomposition of organic matter and mineral nutrient recycling --- in critical ecosystem services. Phenomena such as climate change, seasonal temperature shifts, and anthropogenic inputs also impact network organization, foraging success, and the outcome of multi-species competitive interactions~\cite{ABear2013a,ABear2013b,Boddy2014}.

Because fungi do not have a centralised system to coordinate development, one can posit from the diversity of recognizable network patterns that each fungal species uses a (slightly) different set of local rules to continuously balance investment in growth, transport efficiency, and resilience that collectively maximize the orgnism's long-term global success~\cite{Heaton2015}. 
However, unlike most species, fungi have a highly plastic morphology with few quantifiable traits~\cite{Aguilar-Trigueros2015}. Consequently, most descriptions of fungal behaviour to date have relied on relatively simple growth measures ---
such as hyphal growth rate, branching frequency, or fractal dimension --- coupled with qualitative descriptors of life-history strategy, such as `resource-restricted' or `non-resource-restricted' (depending on whether the vegetative mycelium can use an existing resource as a base to forage over greater distances) or `guerilla' and `phalanx' growth forms (which contrast rapidly growing, sparse networks with dense, highly cross-linked networks)~\cite{Boddy1999, Buss1991}. In some species, these differences in growth morphology are typically correlated with a set of life-history traits; they identify trade-offs in resource allocation, habitat selection, and functional diversity within the community assemblage~\cite{Aguilar-Trigueros2015}. However, unlike in other species, a common framework for phenotypic trait analysis in fungi is only just beginning to emerge~\cite{Aguilar-Trigueros2015}. Given the importance of network architecture in resource discovery and acquisition, control of territory, stress tolerance, and resilience to damage or attack, it is critical (1) to develop quantitative measures that can capture subtle changes in growth form and network organization and (2) to evaluate how phenotypic characteristics contribute to fungal population dynamics, community structure, and ecosystem functioning~\cite{Aguilar-Trigueros2015}.
  
Prior measures of fungal network architecture have characterized population distributions of morphological features in fungal colonies. Such features include both local measures (such as number of tips, node degree at branch points, and branch length~\cite{Fricker2008, Fricker2009, Bebber2007, Barry2009, Vidal-Diez2015}) and global measures (such as fractal dimension~\cite{Boddyfractal1999}, predicted global transport efficiency, and resilience~\cite{Heaton2012, Fricker2008, Fricker2009, Bebber2007}). 
However, most of the phenotypic plasticity and behaviour is likely to arise at an intermediate scale (i.e., a mesoscale) that reflects how smaller units (hyphae, branches, and cords) are organized locally to produce spatial domains with differing architecture and behaviour that collectively yield global behaviour and temporal changes in such behaviour. 
For example, the addition of a new resource causes major strengthening of cords that link to an existing resource, which in turn results in an increase in both resilience and local transport efficiency~\cite{Bebber2007}. Conversely, damage due to grazing initially reduces the integrity of a network but can then stimulate local proliferation of fine hyphae, which then fuse and reconnect, thereby increasing the resilience of the damaged region~\cite{ABear2013a, ABear2013b, Tordoff2006, Rotheray2008, Boddy2010}. 
Consequently, the network formed by each species should not be construed as a single homogeneous structure with uniform architecture and behaviour, but rather as a dynamic and heterogeneous system that operates within some species-specific constraints.  

In many networked systems, there has been progress in characterizing intermediate scales of network organization using various methods of coarse-graining. It is especially prominent to study `community structure', in which one seeks to algorithmically detect densely-connected sets of nodes that are sparsely connected to other densely-connected sets of nodes~\cite{Porter2009, Fortunato2010}. Other mesoscale network features include core--periphery structure \cite{csermely2013}, block models \cite{doreianbook}, and roles and positions \cite{rolesim2014}. Algorithmic detection of mesoscale network structures has the potential to provide objective measures of the subtle intra-specific variation in adaptive traits in fungi (and in other organisms) and to highlight diversity between different species.  Constructing taxonomies of fungal network architecture can thereby provide insights into adaptive fungal behaviour and help elucidate similarities and differences among the underlying rules that govern behaviour. Because a fungus is essentially a living network, we describe changes in fungal network architecture across time as `network behaviour'.

It is also useful to compare fungal networks to other network-forming organisms, such as the acellular slime mould ({\em Physarum polycephalum}), or clonal invertebrates, such as {\em Hydractinia echinata}, which are taxonomically very distinct from fungi but still form extensive reticulate networks with a contiguous lumen (i.e., central cavity) that allows long-distance internal fluid flows and nutrient circulation. {\em Physarum} is essentially a single giant multi-nucleate animal cell that forms extensive tubular networks with dominant, interconnected, many-branched `veins' that form a hierarchy of hydraulically-coupled loops. The {\em Physarum} network is also dynamic. On a short timescale (minutes), actin--myosin contractions drive rapid peristaltic cytoplasmic flows (`shuttle-streaming') to ensure rapid mixing of internal resources through a foraging network. On a longer timescale (hours), some tubes thicken, typically in response to local resource distribution, and tubes that have lower\footnote{The flows are lower both in volume and in speed.} flows regress (see, e.g., Ref.~\cite{Tero2010}). 
Hydractiniid hydroids form colonies that are connected by a common gastrovascular system that consists of feeding polyps and extending stolons that can also fuse to form a stolonal mat~\cite{Blackstone1991, Buss1991}. They also exhibit peristaltic flows, which are driven by contraction of the muscular polyps. 

One can grow fungi, slime moulds, and hydractiniid hydroids in a laboratory, and it is consequently possible to expose them to a wide variety of experimental conditions and species interactions, in multiple replicates, to generate a rich collection of networks. Therefore, investigating such adaptive, self-organized networks --- which are honed by evolution --- provides a fascinating opportunity to uncover underlying principles of biological network organization, evaluate the relevance of network descriptors (which have been developed in related disciplines) to evolved network behaviour, and explore how much utility biologically-inspired algorithms have in other domains~\cite{Fricker2009, Tero2010, Kunita2013}.


\section{Data and methods} \label{sec:data_and_methods}

To make progress in the study of fungal networks, it is important to develop tools to characterize their structure, their function, and how they develop over time and when using different treatments. There is a long history of qualitative description of fungal networks that dates back to the seminal work of Buller in the 1930s (see, e.g., \cite{Buller1931}). More recently, network characterizations have been based on translating a mycelial image to a planar, weighted, undirected graph \cite{Fricker2009, Bebber2007, Heaton2012a}. In such characterizations, the nodes are located at hyphal tips, branch points, and anastomoses (i.e., hyphal fusions). The edges represent cords, and their weights\footnote{Note that \cite{Onnela2012} used the cylindrical volume $V = \pi r^2 L$ for edge weights in fungal networks, although the authors of that paper mistakenly wrote that they used conductance.} are determined from the Euclidean length ($L$) and radius ($r$) of each cord combined either as (1) the cylindrical volume $V = \pi r^2 L$ to represent the biological `cost' of the cord or (2) the predicted conductance $G = r^2/L$. The conductance assumes that the cords are bundles of equal-sized vessels, so the aggregate conductance scales with the cross-sectional area of the whole cord~\cite{Bebber2007}. By contrast, for a single vessel, the conductance would scale with $r^4$ for Pouiseille flow.

In the present paper, we refer to the above network representations as `structural networks'. Simple network measures, such as notions of `meshedness' (for planar networks), clustering coefficients, and betweenness centrality, have been calculated from graph representations of fungal networks \cite{Bebber2007}. However, the computation of simple diagnostics has not been able to capture the subtle differences in spatial structure between species or in the same species when they are responding to different experimental conditions~\cite{Heaton2012a}. Although such features are hard to describe quantitatively, human observers are able to see them qualitatively, and it is desirable to develop effective ways to also capture them quantitatively.

We compare the structural networks, for which we use only the predicted conductance $G$ of each cord, to one that is based on a predicted functional view of the importance of each cord for transport.  For the latter (our so-called `functional networks'), we calculate the weight of each cord using a `path score', which is a diagnostic (see the definition below) that measures the importance of an edge for transport of nutrients in a network in a way that is more nuanced than standard measures of betweenness centrality~\cite{SHLee2014}. Such a relation between the structural properties of networks and dynamics on networks is crucial in other systems as well (e.g., in ecological systems~\cite{Scotti2012}).
  
The computation of path scores also highlights core--periphery structures that are based on transport properties rather than on the usual density-based notions of such structures~\cite{csermely2013}. In a fungal (or slime-mould) network, we expect core cords to highlight dense parts of the network near the inoculum (i.e., source material for a new culture) or in parts that connect to additional resources, whereas the periphery parts of a network can correspond to the foraging margin. Transport-based measures of core--periphery structure for both nodes and edges in networks were investigated recently in a wide variety of networks and using different transportation strategies (e.g., both geodesics and random walks)~\cite{SHLee2014}.  (See~\cite{mihai2014} for related theoretical work.) Because one of the primary predicted functions of fungal networks is nutrient transport, it is more appropriate to examine the `coreness' properties of edges (i.e., cords) rather than nodes.
  
  As discussed in~\cite{SHLee2014}, we quantify a transport-based measure of coreness called the \emph{path score} (PS) for each edge by examining which cords appear most often on `backup paths' if any particular cord is broken. This measure thereby incorporates elements of both betweenness centrality and network resilience. We expect that core edges in a network should occur more frequently than peripheral edges in short backup paths.  One can define path scores for both directed and undirected networks and for both weighted and unweighted networks.  We treat the networks that we construct from our fungal systems as weighted and undirected.
  
We denote the set of edges by $\mathbb{E} = \{ (j,k) | $ node $j$ is adjacent to node $k \}$ and the number of edges by $M = | \mathbb{E} |$. The PS for edge $e$ is defined by
  \begin{equation}
  	\textrm{PS}(e) = \frac{\displaystyle 1}{M} \sum_{(j,k) \in \mathbb{E}} \sum_{\{ p_{jk} \}} \sigma_{jek} [ \mathbb{E} \setminus (j,k) ]\,,
  \label{eq:PS_formula}
  \end{equation}
  where $\sigma_{jek} [ \mathbb{E} \setminus (j,k) ] = 1/|\{ p_{jk} \}|$ if edge $e$ is in the set $\{ p_{jk} \}$ that consists of `optimal backup paths' from node $j$ to node $k$ (where we stress that {the edge} $(j,k)$ {is removed from} $\mathbb{E}$) and $\sigma_{jek} [ \mathbb{E} \setminus (j,k) ] = 0$ otherwise. A backup path is an alternative path from the same source to the same target when the direct connection from the source to the target is broken. An optimal backup path is a shortest path among those backup paths. We remark in passing that a notion of coreness based on response to node removal was used in \cite{Valente2010} in the context of closeness centrality (rather than betweenness centrality).
  
To determine an optimal path between nodes $i$ and $j$, we find the backup path $p_b (i;j)$ that consists of the set of connected edges between $i$ and $j$ that minimizes the sum $\sum_{(k,l) \in p_b(i;j)} R_{kl}$ of the resistances $R_{kl}$ for all edges $(k,l)$ of a network in which the edge $e_{ij} := (i,j)$ has been removed.  The resistance of edge $(k,l)$ is given by $R_{kl} = 1/G_{kl}$, where the conductance is $G_{kl} = r^2 / L$, the quantity $r$ is the radius of a cord, and $L$ is the length of edge $(k,l)$. We set $R_{kl} = 0$ (instead of $R_{kl} = \infty$) when an edge is removed because the edge simply does not exist. To capture a functional view of the fungal networks (i.e., to obtain so-called `functional networks'), we also construct weighted networks in which we preserve topology but use PS values instead of conductance values as the edge weights.

Detailed measurements and modelling have been used to experimentally examine the development of network architecture over time, predict the flow of water and nutrients through the resulting empirical network using an advection and diffusion model, and compare model output with experimentally measured radiolabel distributions used to track actual nutrient movement~\cite{Heaton2012a, Heaton2010}. This approach has revealed good correlations between growth-induced fluid flows and nutrient transport, and one can even see hints of the local rules that optimize behaviour, but it is too technically demanding and costly to be used as a routine pipeline for analysis of network behaviour across a multitude of data sets.  Other approaches are thus necessary to compare the structures and functions of a large set of fungal species, or to compare the same species over time or in different experimental conditions.
  
In a recent paper~\cite{Onnela2012}, two of us (and our coauthors) illustrated that examining community structure of fungal networks using mesoscopic response functions (MRF) provides biologically-sensible clusterings of different species and developmental stages for a particular species. The MRFs in \cite{Onnela2012} were calculated by maximizing the objective function `modularity' \cite{Newman2006,newman2010,Porter2009,Fortunato2010} for numerous values of a resolution parameter and calculating how some quantities change as a function of the resolution parameter. We thereby have a `response curve' for how a given quantity changes as one sweeps across resolution-parameter values. The authors of \cite{Onnela2012} measured MRFs for normalized versions of three quantities: maximum modularity (to quantify how well a network can be partitioned into communities), entropy (to quantify the heterogeneity of communities sizes), and number of communities. For any type of MRF, one can then compare the MRFs for any two networks to calculate a distance between those networks. This yields a distance between each pair of networks in a collection. 

In the present paper, we explore the utility of an MRF-based classification using a large set of fungal networks (with many more than those studied in \cite{Onnela2012}) that includes a wide variety of different developmental stages, nutrient regimes, growth substrates, competition, and levels of predation (see Table~\ref{network_code}).  As an accompaniment to our paper, we also make a large data set of fungal networks available in the public domain. (See the Supplementary data in Sec.~\ref{sec:supplementry_data}.)

  \begin{table*}
  	\caption{Species and experimental conditions for the fungal networks in Fig.~\ref{MRF_total_levels}. (A `level' in the table represents a discrete valuation that ranges from the minimum to the maximum of the associated quantity.)
  	}
  	\centering
  	\begin{tabular}{lll}
  		\hline
  		\hline
  		Attribute &  Code/Level (Colour) &  Descriptions \\ 
  		\hline
  		Species &  \emph{Pp} &  \emph{Physarum polycephalum}: an acellular slime mould  \\
  		&   &  that forms networks but is taxonomically  \\
  		&   & distinct from fungi \\
  		\cline{2-3}
  		&  \emph{Pv} &  \emph{Phanerochaete velutina}: a foraging saprotrophic  \\
  		&   &  woodland fungus that forms reasonably  \\
  		&   & dense networks \\
  		\cline{2-3}
  		&  \emph{Ag} &  \emph{Agrocybe gibberosa}: a foraging saprotrophic fungus   \\
  		&   &  that is isolated from garden compost and \\
  		&   & forms dense networks \\
  		\cline{2-3}
  		&  \emph{Pi} &  \emph{Phallus impudicus}: forms regular, highly cross-linked  \\
  		&   & networks but grows relatively slowly \\
  		\cline{2-3}
  		&  \emph{Rb} &  \emph{Resinicium bicolor}: forages rapidly with a sparse \\
  		&   & network that is not very cross-linked \\
  		\cline{2-3}
  		&  \emph{Sc} &  \emph{Strophularia caerulea}: a foraging saprotrophic  \\
  		&   & woodland fungus that is isolated from birch   \\
  		&   & woodland and forms dense networks \\
  		\hline
  		resources &  I/min (blue) &  Initial colonised wood block (inoculum; I) only~\cite{Bebber2007, Heaton2010, Heaton2012}\\
  		\cline{2-3}
  		&  I+R/low &  Inoculum plus a single additional wood-block resource (R)~\cite{Bebber2007}. \\
		&   & For display purposes, we group the increasing amounts of \\
		&   & resource into 5 discrete levels. \\
  		\cline{2-3}
  		&  I+4 $\times$ R/intermediate &  Inoculum plus four additional wood-block resources \\
  		&   & (positioned as a cross) ~\cite{Boddy2010}\\
  		\cline{2-3}
  		&  I+4 $\times$ R/intermediate &  Inoculum plus four wood-block resources placed  \\
  		&   & together ~\cite{Fricker2008}\\
  		\cline{2-3}
  		&  5 $\times$ I/level 3 &  Five inocula placed in a pentagonal arrangement \\
                &                       & (D. Leach and M.D. Fricker, unpub.) \\
  		\cline{2-3}
  		&  Tokyo/level 4 &  Pattern of oat flakes placed to match the major cities \\
  		&   &  around Tokyo ~\cite{Tero2010}\\
  		\cline{2-3}
  		&  UK/max (red) &  Pattern of oat flakes placed to match the major cities  \\
  		&   & in the UK (S. Kala and M.D. Fricker, unpub.)\\
  		\hline
  		grazing &  U/min (blue) &  Ungrazed \\
  		\cline{2-3}
  		&  \emph{Fc}/low &  \emph{Folsomia candida}: a small soil arthropod that grazes  \\
  		&  & on fungal networks with low density  \\
  		&  & (10 per microcosm)~\cite{Tordoff2006} \\
  		\cline{2-3}
  		&  \emph{Fc} or \emph{Fc}-M/intermediate &  \emph{Fc} with medium density  \\
  		&  & (20 per microcosm)~\cite{Tordoff2006, Rotheray2008} \\
  		\cline{2-3}
  		&  \emph{Fc}-H/max (red) &  \emph{Fc} with high density (40 per microcosm)~\cite{Tordoff2006} \\
  		\hline
  		substrate &  A/min &  Agar: used as a growth medium (substrate) for \\
  		&   & \emph{Physarum polycephalum} ~\cite{Tero2010}\\
  		\cline{2-3}
  		&  B/intermediate &  Black sand: a nutrient-free substrate used for \\
  		&   & radiolabel-imaging experiments ~\cite{Fricker2008, Heaton2012a} \\
  		\cline{2-3}
  		&  S/max &  Compressed, non-sterile soil that closely  \\
  		&   & represents the natural growth environment  \\
  		&   & for the fungi ~\cite{Boddy1999}\\
  		\hline
  		interaction & N/min &  No interaction: fungal species grown on its own \\
  		\cline{2-3}
  		&  Hf/max &  Grown in competition with \emph{Hypholoma fasciculare} \\
  		\hline
  	\end{tabular}
  	\label{network_code}
  \end{table*}
  

\section{Results}
\label{sec:results}

\begin{figure*}
\centering
\includegraphics[width=0.70\textwidth]{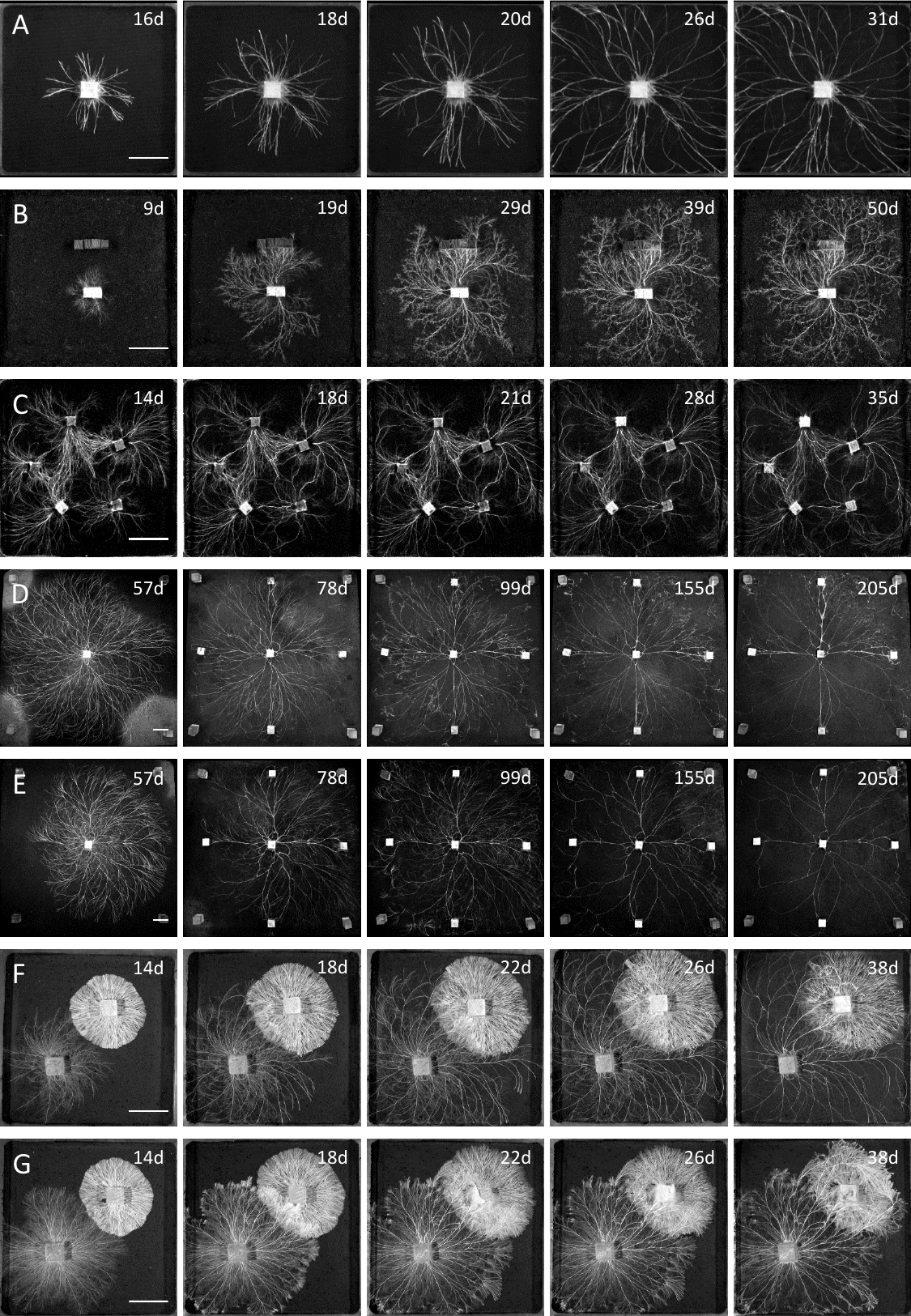}
\caption{
\textsf{(A)} Growth of {\em Resinicium bicolor} on soil (in a 220 mm $\times$ 220 mm microcosm) as an example of a relatively sparse network.
\textsf{(B)} Growth of {\em Phanerochaete velutina} on black sand with an additional set of four wood-block resources.
\textsf{(C)} Network formation in {\em Phanerochaete velutina} from day 14 to day 35 on a compressed black-sand substrate from a pentagonal arrangement of wood-block inocula.
\textsf{(D)} Large microcosm (570 mm $\times$ 570 mm) of {\em Phanerochaete velutina} supplemented with four additional resources. The network begins to regress as it consumes the resources.
\textsf{(E)} Similar experimental microscosm to (D), except that grazing insects were added on day 49.
\textsf{(F)} {\em Phanerochaete velutina} growing in competition with {\em Hypholoma fasciculare}.
\textsf{(G)} {\em Phanerochaete velutina} growing in competition with {\em Hypholoma fasciculare} in the presence of grazing insects.
Each scale bar (see the left panels) represents 50 mm, and the upper right corner of each panel gives the amount of growth time in days. The images in row \textsf{(A)} are from \protect\cite{Tordoff2006}; those in row \textsf{(B)} are from \protect\cite{Fricker2008}; those in row \textsf{(C)} are from D. Leach and M.\,D. Fricker, unpublished; those in rows \textsf{(D)} and \textsf{(E)} are from \protect\cite{Boddy2010}; and the images in rows \textsf{(F)} and \textsf{(G)} are from \protect\cite{Rotheray2008}.}
\label{timeseries_figure}
\end{figure*}

We present several examples in Fig.~\ref{timeseries_figure} to give a visual indication of the challenges that face biologists when trying to describe variation in network organization across different species and under different experimental treatments, as the structural differences in the various scenarios can be rather subtle. We show time series of fungal growth for one species ({\em Resinicium bicolor}) that tends to grow as a relatively sparse network [see row \textsf{(A)}; data and images from~\cite{Tordoff2006}] and a second species ({\em Phanerochaete velutina}) that forms more cross-links [see row \textsf{(B)}; data and images from~\cite{Fricker2008}]. For the latter, we illustrate the impact of increasingly complex microcosms when there is variation in both the level and positioning of resources [compare rows \textsf{(B)} and \textsf{(C)}; the latter set of images and data are from D. Leach and M.\,D. Fricker, unpublished], resources become depleted and the networks shrink in both the presence and the absence of attack by mycophagous insects [see rows \textsf{(D)} and \textsf{(E)}; data and images from~\cite{Boddy2010}], and networks grown in competition with another species ({\em Hypholoma fasciculare}) both with and without predation [see rows \textsf{(F)} and \textsf{(G)}; data and images from~\cite{Rotheray2008}].

In general, the presence of additional resources leads to the strengthening of connections between an initial inoculum and a new resource [see rows \textsf{(B)}--\textsf{(G)} of Fig.~\ref{timeseries_figure}], and these cords are more persistent and resilient to damage than the remainder of the colony under grazing pressure [see row \textsf{(E)}] or during interspecies combat [see rows \textsf{(F)} and \textsf{(G)}]. In the latter case, colony growth also becomes highly asymmetric as the {\em Phanerochaete} overgrows the {\em Hyphaloma} colony and captures its resource base. Neither microscopic measures (such as node degree) nor macroscopic measures (such as network diameter) quantitatively capture such changes in network architecture, although these changes are visible to human observers.

\begin{figure*}
	\centering
	\includegraphics[width=0.70\textwidth]{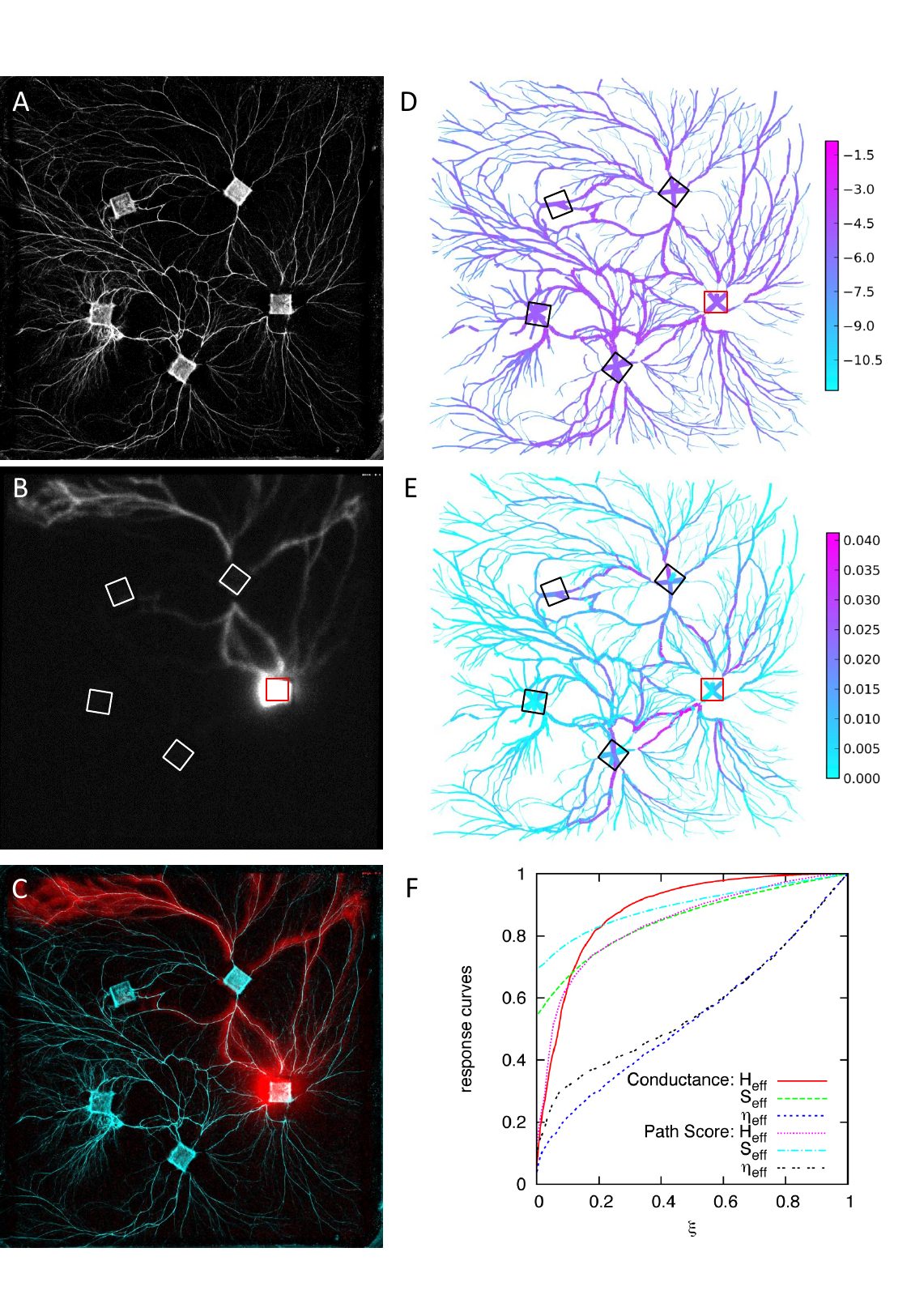}
	\caption{\textsf{(A)} One of the fungal networks formed by {\em Phanerochaete velutina} after 30 days of growth on a compressed black-sand substrate from a pentagonal arrangement of wood-block inocula. \textsf{(B)} Path of radiolabeled nutrient ($^{14}$C-amino-isobutyrate) added on day 30 and imaged using photon-counting scintillation imaging (PCSI) for 12 hours. \textsf{(C)} Merged overlay of panels \textsf{(A)} and \textsf{(B)} to highlight the path that is followed by the radiolabel. \textsf{(D)} We colour the edges of the manually digitized network according to the logarithm of the conductance values. Edge thickness represents cord thickness. \textsf{(E)} We colour the edges according to the path score (PS) values of the fungal network. \textsf{(F)} MRF curves for conductance-based and PS-based weights.  We show MRF curves for the effective energy $H_{\mathrm{eff}}$, effective entropy $S_{\mathrm{eff}}$, and effective number $\eta_{\mathrm{eff}}$ of communities as a function of the resolution parameter $\xi$.  See \cite{Onnela2012} for details on MRFs, and note that the effective energy is a rescaled version of the negative of optimized modularity. For the MRF analysis, we remove nodes with degree $k=2$, and we adjust the weights of the edges that connect the remaining nodes to include the values for each $k=2$ segment by treating each of these edges as a set of resistors in series. (The edges in panels \textsf{(D)} and \textsf{(E)} include nodes with degree 2, as they are needed to trace the curvature of the cords.)	
	}
	\label{Pv_PS_figure}
\end{figure*}

In addition to examining network architecture, we are also interested in the function of fungal networks with respect to long-distance nutrient transport from sources (wood blocks) to sinks (growing hyphae at the foraging margin). For example, in Fig.~\ref{Pv_PS_figure}, we show a network formed by {\em Phanerochaete velutina} growing from five wood-block inocula that are placed in a pentagonal arrangement on a compressed black-sand substrate. [It is a similar arrangement as in Fig.~\ref{timeseries_figure}\textsf{(C)}.] The network emerges [see Fig.~\ref{Pv_PS_figure}\textsf{(A)}] as cords fuse and are strengthened, or as they are recycled and disappear. The fungal network that forms has a relatively densely interconnected core and relatively tree-like foraging branches on the periphery. One can map functional flows in the network with radiotracers that use photon-counting scintillation imaging (PCSI) to provide a snapshot of nutrient transport [see panel \textsf{(B)} and the overlay in Fig.~\ref{Pv_PS_figure}\textsf{(C)}] at a particular time instant in the same network. In this experiment, a non-metabolized amino-acid ($^{14}$C-amino iso-butyrate; AIB) was added to one wood block inoculum. [We show the inoculum using a dark square (red online) in Fig.~\ref{Pv_PS_figure}\textsf{(B)} and using a solid square in Figs.~\ref{Pv_PS_figure}\textsf{(D)} and \ref{Pv_PS_figure}\textsf{(E)}.] It was then transported rapidly towards one of the adjacent wood blocks and subsequently to the colony margin. One can extract the network architecture using image processing, and one assigns a conductance to the edge weights based on their length and cross-sectional area~\cite{Heaton2012, Bebber2007,  Onnela2012} to give a `structural network' [see panel \textsf{(D)}] or according to `path score' (PS), which quantifies edge importance~\cite{SHLee2014}, to estimate a `functional network' [see panel \textsf{(E)}]. (See Sec.~\ref{sec:data_and_methods} for detailed definitions of the two types of networks.)  The PS values on the edges of a fungal network reflect movement paths in the region of the colony in which radiolabel was translocated, suggesting that the PS values capture some aspects of real nutrient movement in fungal networks. However, there is not a simple correspondence between PS values and observed nutrient transport, as there are cords with high PS values that could have been utilized to reach the neighbouring wood block towards the lower left even though there was no detectable radiolabel translocation over the 12-hour time period of the measurement.

\begin{figure*}
	\begin{tabular}{l}
		\textsf{A} \\
		\includegraphics[width=\textwidth]{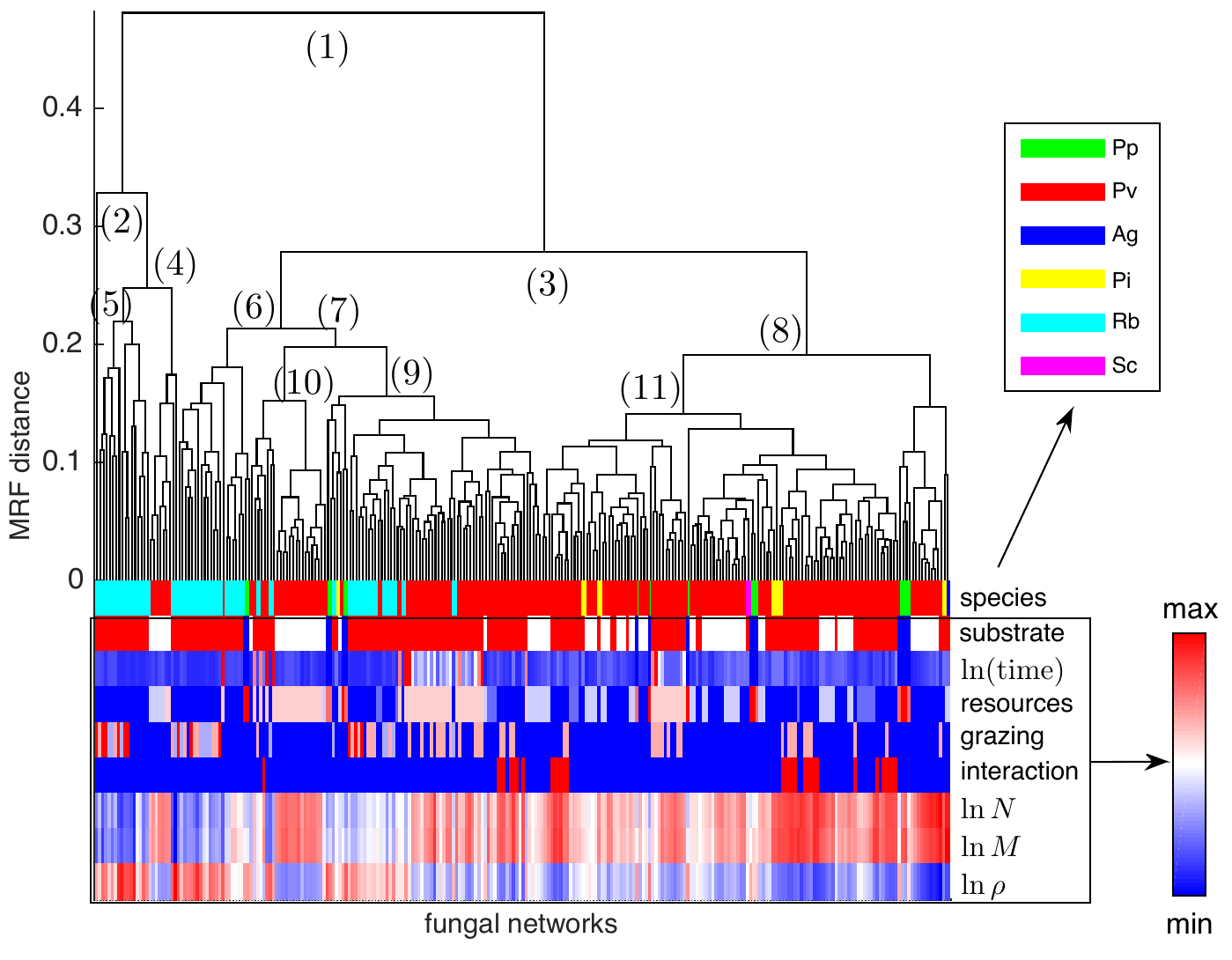} \\
	\end{tabular}
	\caption{Taxonomies of $270$ fungal (and slime-mould) networks determined using \textsf{(A)} conductance $G$ and \textsf{(B)} (next page) PS values~\cite{SHLee2014} as the edge weights. That is, panel \textsf{(A)} gives a structural taxonomy, and panel \textsf{(B)} gives a functional taxonomy. We produce the dendrograms that represent the two taxonomies using an MRF analysis~\cite{Onnela2012}, where we apply average linkage clustering \cite{dataclust2007} to the MRF distance from principal component analysis of three different MRFs (effective energy, effective energy, and effective number of communities)~\cite{Onnela2012}. We use the same methodology (including the determination of community structure using modularity optimization with a resolution parameter) as in \cite{Onnela2012}. See Table~\ref{network_code} for the species abbreviations and the levels of substrate, resources, and grazing; and see the main text in Sec.~\ref{sec:results} for a discussion of the numbered branching points in the dendrograms.		
At the bottom of the taxonomies, we show the logarithms of the number $N$ of nodes, the number $M$ of edges, and the edge density $\rho = 2M/[N(N-1)]$. We label the main branch points in each dendrogram in parentheses.  (Note that `branches' in a fungal network are different from `branches' in a taxonomy.  It is standard to use such terminology in both contexts.)
		}
	\label{MRF_total_levels}
\end{figure*}

Nevertheless, in this example, it is clear that nutrient transport can occur both towards and away from a wood block at the same time, suggesting that nutrient flow does not have a consistent preferred direction. This has been observed in other studies, which explicitly showed that indicate that tracer movement can be bidirectional ~\cite{Lindahl2001, Tlalka2007}. We therefore regard each cord as potentially allowing movement in either direction, so the resulting network is undirected. For each structural and functional network, we generate three MRFs [see panel \textsf{(F)}]~\cite{Onnela2012} to examine network community structure at multiple scales. In Figs.~\ref{MRF_total_levels}\textsf{(A)} and \ref{MRF_total_levels}\textsf{(B)}, we show the resulting taxonomy for $270$ structural [see panel \textsf{(A)}] and functional [see panel \textsf{(B)}] fungal networks. For more details on data and calculations, see Sec.~\ref{sec:data_and_methods}. We also include the data for all networks as Supplementary data, which we describe in Sec.~\ref{sec:supplementry_data}.

\begin{figure*}
	\begin{tabular}{l}
		\textsf{B} \\
		\includegraphics[width=\textwidth]{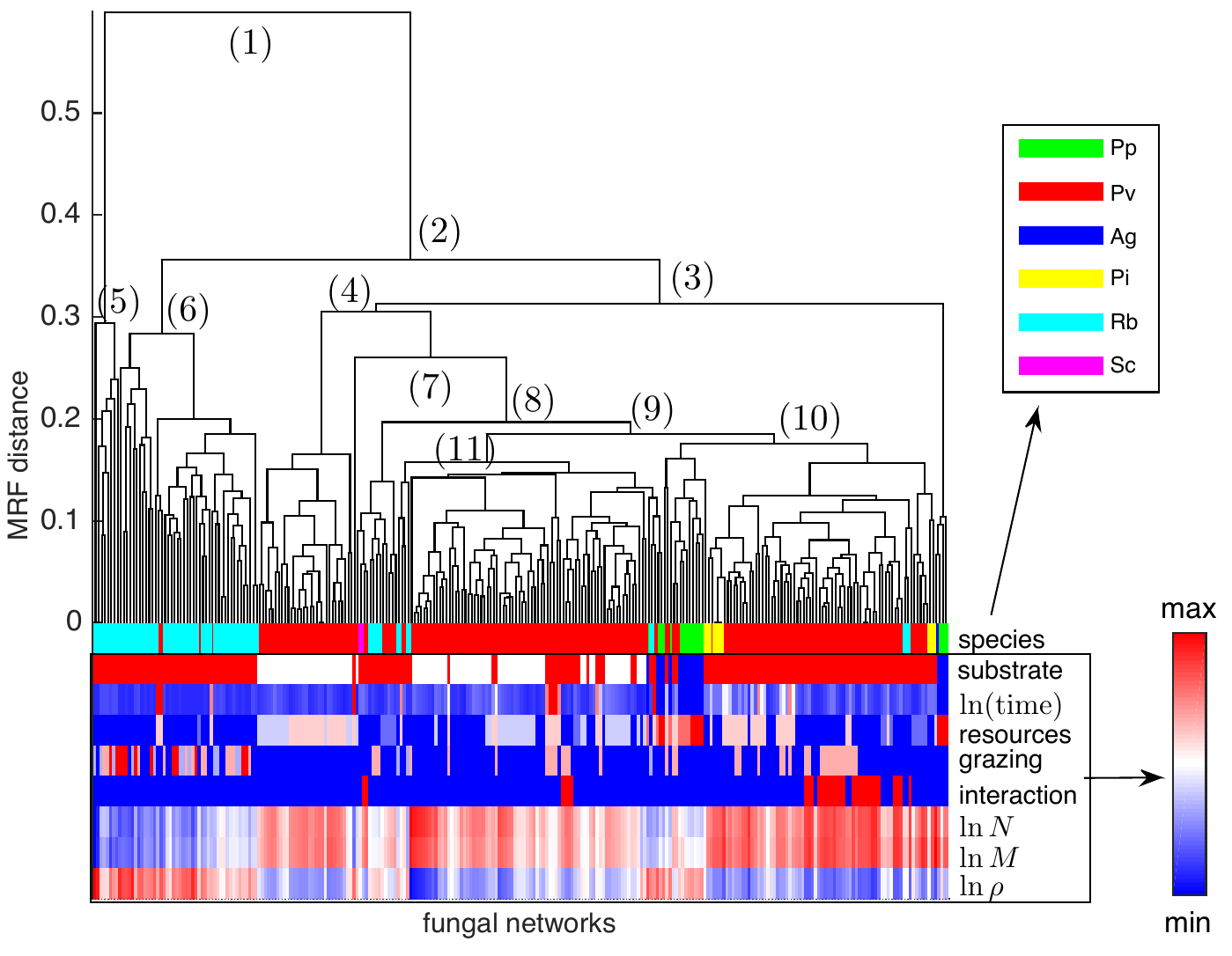} \\
	\end{tabular}
\end{figure*}

In both the structural and functional networks, the diverse selection of experimental conditions is reduced to a set of intuitively sensible clusters. The `functional' PS measure provides more harmonious groupings --- which are clustered by species, substrate, resource level, grazing, and interaction --- than in the structural networks. We also observe that networks that arise from some treatments are spread across the taxonomy. In particular, as large networks of {\em Phanerochaete velutina} deplete their resources, they move from clusters with well cross-linked networks to very sparse networks, similar to the normal growth pattern of {\em Resinicium bicolor}. In Fig.~\ref{MRF_total_levels}, we show taxonomies, in the form of dendrograms, for $270$ fungal networks based on both structure and function. Recall that the structural and functional networks have the same topologies, but their edge weights are different: the weights are given by estimated conductance values for the structural networks and by PS values for the functional networks.

For both the structural and functional fungal networks, the simplest network measures for each leaf (e.g., number of nodes, number of edges, and node density) only reveal a limited correlation with the major branches in the dendrogram.  This suggests that the classification is not trivially dominated by the size of each network and also that it is necessary to go beyond the computation of only such simple measures to produce a reasonable taxonomy. When we color leaves according to the values of the major attributes in each experiment (species, substrate, time point, resource level, and grazing intensity), we observe that groups of species with similar attributes begin to emerge and are visible as substantial contiguous blocks in the dendrograms. Nevertheless, we also observe that each attribute is not uniquely associated with one group; this again suggests that the classifications are not based on a trivial separation by any one of these attributes (e.g., species) alone. This, in turn, suggests that they also reflect similarities in the topologies and edge weights (i.e., geometries) of the networks.

The Pearson correlation coefficient between the MRF distance values (see Appendix B.2 of \cite{Onnela2012}) for the sets of structural and functional network sets is 0.418. (The $p$-value is less than $10^{-308}$, which is the minimum value of floating-point variables in \textsc{Python}.) By contrast, the mean correlation coefficient from $100$ uniform-at-random permutations of the MRF distance values is $2.12 \times 10^{-5}$, with a standard deviation of $3.67 \times 10^{-3}$. We infer that there is some degree of correlation between the weights in the structural and functional networks, although they clearly capture different properties of the fungi.

A key challenge is to try to interpret the taxonomic groupings from a biological perspective to obtain insights that cannot be captured from qualitative descriptions of each network, particularly when making comparisons between different experiments from different laboratories over an extended time period. To do this, we follow the major branch points of the dendrograms in a top-down exploration of each taxonomy.  We label branches in the dendrograms in the order in which they occur in the taxonomic hierarchies. In the conductance-based (`structural') classification [see Fig.~3\textsf{(A)}], a small group splits off at a high level (branches 1, 2). This group then separates into two parts: one contains \emph{Resinicium bicolor} (\emph{Rb}) with some grazing (5), and the other contains \emph{Phanerochaete velutina} (\emph{Pv}) grown on black sand (4). The other main branch splits to give two clusters (3), but the underlying rationale is not immediately obvious, as both parts include a mixture of different conditions of the attributes (see Table~\ref{network_code}). The clearest subsequent groupings emerge as clusters of \emph{Rb} with grazing at earlier time points (6, 9) and \emph{Pv} on black sand with high resources (10).

Following the same top-down approach, the PS-based (`functional') taxonomy [see Fig.~3\textsf{(B)}] gives groupings that are easier to interpret than the ones from the structural taxonomy. The first set of high-level branch points (1, 2, 5, and 6) all separate clades of \emph{Rb}, and subsequent divisions reflect the level of grazing.
This division seems to capture the functional behaviour of {\em Resinicium} that is typically ascribed to a `guerilla' strategy with rapid but relatively sparse exploration. Branch point (4) separates a group of \emph{Pv} on black sand with relatively high levels of resource and demonstrates that intra-specific variation arising from growth of the same species in different regimes can be identified from subtle variation in the PS-based MRF. Branch point (7) yields a single \emph{Pv} network from one of the large, shrinking network sequences. Interestingly, these shrinking networks are interspersed across the whole dendrogram. (See the isolated pink and red bars in the `$\ln(\mathrm{time})$' row.) During development, these large networks initially cluster with other well-connected networks, but they progressively shift towards clustering better with sparser networks as they regress until they eventually cluster with the \emph{Rb} networks. 

The other arm of branch point (7) leads to a large grouping that contains a set of well-defined clusters. Branch point (8) splits off from a small group that has both \emph{Rb} and \emph{Pv}, but there is no clear common linkage. Branch point (9) yields a large group that is composed predominantly of \emph{Pv} on black sand (with subgroups based on resource levels) and a few interspersed large networks, followed by a well-defined set of groups that lie under branch point (10). The first of these clusters contains most of the \emph{Pi} and \emph{Pp} networks (although a few such networks are located in the nearby clusters), and the second cluster has sequential groups of \emph{Pv} with high levels of resource but little grazing, a group with both grazing and species interactions, and a group with only species interactions.


\section{Discussion}
\label{sec:discussion}

To compare the properties of the various structural and functional networks, we produced taxonomies of the fungal networks from MRFs of each network \cite{Onnela2012} to highlight mesoscale structures from communities~\cite{Porter2009, Fortunato2010}. In network terms, communities are meant to represent nodes that are densely connected internally and which have sparse connections to other communities (relative to a null model). To identify community structure, we optimize a multiresolution version of the modularity objective function. We use the Newman--Girvan null model augmented by a resolution parameter and examine communities of different size scales by tuning the resolution parameter~\cite{newman2010,Reichardt2006}. For each network, we obtain curves for several scaled quantities (effective number of communities, effective energy, and effective entropy) as a function of the resolution parameter. These diagnostics yield a mesoscale fingerprint for each network. Two networks are close to each other in a taxonomy if their MRF curves have similar shapes to each other. (See \cite{Onnela2012} for details.)  Our two taxonomies---one for the structural networks and another for the functional networks---give a pair of `family trees' that describe how closely the various networks are related in the form of a dendrogram.
In general terms, species are well separated by both classifications, with the functional taxonomy also giving groups with more consistent membership from different treatments.
This confirms the view that it is possible to use mesoscale properties to make quantitative estimates that capture intra-specific plasticity in network architecture. There is also overlap between the different treatments, with networks from some time-series experiments spread across several clusters. It is not surprising that the structural and functional taxonomies both contain fine-grained complexity in their terminal groups, as several of the attributes have opposing effects that depend on both the developmental age of each species and the combination of treatments. For example, as a fungal network grows, it tends to change from a branching tree to a more highly cross-linked network through hyphal and cord fusions that connect to each other. The core parts of a fungal network subsequently start to thin out as it explores further until resources run out; the network progressively recycles more cords and again becomes a very sparse network~\cite{Bebber2007,Boddy2010}.

Some of the clearest clusters in the functional taxonomy correlate with substrate, as there are distinct branches in the taxonomy that consist predominantly of \emph{Pv} grown on black sand. Consequently, even though \emph{Pv} is well-represented in the dendrogram, there is a distinguishable effect of substrate on network architecture that is not immediately obvious to a human observer. Likewise, it is fascinating that the functional taxonomy includes clear signatures that correlate with resource level, grazing, and interactions with other species. Such observations underscore the fact that taxonomical groupings of fungal networks that are derived through network analysis can be of considerable assistance to biologists in their attempts to capture the impact of treatment combinations on network behaviour. The construction and analysis of network taxonomies also allow objective groupings of networks across species, treatments, and laboratory settings.

Constructing structural and functional taxonomies has the potential to be important for the development of increased understanding of subtle behavioural traits in biological networks. This type of approach should become progessively more useful as more networks are included in a classification --- particularly if at least some have associated experimentally-validated functional attributes \cite{Heaton2012a,Heaton2010}. Recently developed, sophisticated network extraction algorithms~\cite{Obara2012} can dramatically improve the speed, accuracy, and level of detail of fungal networks. They also facilitate automated, high-throughput processing of fungal network images, which can in turn be used to construct a richly detailed set of networks that are ripe for study via structural and functional network taxonomies.


\section{Conclusions}
\label{sec:conclusions}

We calculated MRFs for a large set of networks of fungi and slime moulds.  We considered two types of networks: (1) `structural' networks in which we calculate edge weights based on conductance values and (2) `functional' networks in which we calculate edge weights based on an estimate of how important edges are for the transport of nutrients.  Calculating MRFs for the fungal and slime-mould networks in each of these two situations makes it possible to construct taxonomies and thereby compare large sets of fungal networks to each other. We illustrated that network taxonomies allow objective groupings of networks across species, treatments, and laboratories. Such classification provides fine-grained structure that indicates subtle interplay between species, substrate, resource level, grazing pressure, and inter-species competition. We also observed that networks that undergo major transitions, such as regressing from a densely connected meshwork to a sparse tree as resources run out, are dispersed across the taxonomies. This reflects a change in their functional behaviour amidst such transitions. 
We propose that measurements of mesoscale structure provide a viable route to quantify highly plastic development and behaviour in fungal networks and that they can provide useful inputs to developing trait-based understanding of fungal ecology~\cite{Aguilar-Trigueros2015}.  





\section{Supplementary data}
\label{sec:supplementry_data}

Supplementary data are available online at
\url{http://newton.kias.re.kr/~lshlj82/fungal_networks_MATLAB.zip}.
The data include the sparse adjacency matrices (denoted by
\textsf{A}) and the coordinate matrices (denoted as
\textsf{coordinates} of the network nodes) for each of the fungal networks in \textsc{Matlab} format. For the coordinates, the first and second columns, respectively, represent horizontal and vertical coordinates.  We use the terminology in Table~\ref{network_code} and Sec.~\ref{sec:data_and_methods} to name
the files. The folders \texttt{Conductance} and
\texttt{PathScore}, respectively, contain the conductance-based
(`structural') and PS-based (`functional') edge weights. 
We also provide the complete list of fungal networks as a
spreadsheet file (\texttt{list\_of\_fungal\_networks.xlsx}) in
Microsoft \textsc{Excel} format.


\section*{Acknowledgements}

SHL and MAP acknowledge a grant from the EPSRC [EP/J001759/1]. SHL was also supported by Basic Science Research Program through the National Research Foundation of Korea (NRF) funded by the Ministry of Education [2013R1A1A2011947]. MDF acknowledges grant RGP0053/2012 from the Human Frontier Science Program. We thank M.~Anantharaman, S.~Kala, D.~Leach, G.~Tordoff for collecting images and networks that have not yet been published. We also thank U. Brandes and anonymous referees for helpful comments. We used MRF source code that is available at \url{http://www.jponnela.com/web_documents/mrf_code.zip}.


%



\end{document}